# Asymmetrically Encapsulated vertical ITO/MoS$_2$/Cu$_2$O photodetector with ultra-high sensitivity

Sangeeth Kallatt[1,2], Smitha Nair[2], and Kausik Majumdar[1*]

[1]Department of Electrical Communication Engineering, Indian Institute of Science, Bangalore 560012, India

[2]Center for Nano Science and Engineering, Indian Institute of Science, Bangalore 560012, India

*Corresponding author, email: kausikm@iisc.ac.in

**ABSTRACT:** Strong light absorption, coupled with moderate carrier transport properties, makes two-dimensional (2-D) layered transition metal dichalcogenide (TMD) semiconductors promising candidates for low intensity photodetection applications. However, the performance of these devices is severely bottlenecked by slow response with persistent photocurrent due to long lived charge trapping, and nonreliable characteristics due to undesirable ambience and substrate effects. Here we demonstrate ultra-high specific detectivity ($D^*$) of $3.2 \times 10^{14}$ Jones and responsivity ($R$) of $5.77 \times 10^4$ AW$^{-1}$ at an optical power density ($P_{op}$) of 0.26 Wm$^{-2}$ and external bias ($V_{ext}$) of $-0.5$ V in an indium tin oxide (ITO)/MoS$_2$/copper oxide (Cu$_2$O)/Au vertical multi-heterojunction photodetector exhibiting small carrier transit time. The active MoS$_2$ layer being encapsulated by carrier collection layers allows us to achieve negligible trap assisted persistent photocurrent and repeatable characteristics over large number of cycles. We also achieved a large $D^* > 10^{14}$ Jones at zero external bias due to the built-in field of the asymmetric photodetector. Benchmarking the performance against existing reports in literature shows a pathway for achieving reliable and highly sensitive photodetectors for ultra-low intensity photodetection applications.

# 1. Introduction

Ultra-low intensity photodetection is of significant technological importance in a variety of fields covering medical instrumentation, remote sensing, strategic sector, space, and industrial applications. Over the past few years, layered 2-D materials, including TMD semiconductors, have been explored extensively for electronic and optoelectronic applications[1–7]. One of the key advantages of layered TMD materials is that they do not require any stringent epitaxial growth and high quality crystals can be deposited over any arbitrary substrate[8,9]. Such low-cost material deposition, coupled with excellent light absorption and moderate carrier transport properties, makes these layered TMD materials promising candidates for sensitive and inexpensive photodetection applications. While large responsivity has been demonstrated in layered material based photodetectors[10–33], in general, there are two important bottlenecks that limit the practical applicability of these devices. First, the response of many of the 2-D devices is slow owing to long lived charge trapping in the active material as well as in the substrate. While such long lived traps do provide enormous gain in the 2-D photoconducting devices [17,28,34,35], they tend to slow down the photoresponse of the device with appreciable persistent photocurrent. Second, the very nature of the two-dimension exposes the active part of these devices to the surrounding, leading to unreliable and poorly repeatable characteristics due to ambience and substrate induced detrimental effects.

In this work, we encapsulate multi-layer $MoS_2$ film between two conducting metal oxides, namely ITO and $Cu_2O$ in a vertical stack. The metal oxide layers, on one hand, protect the $MoS_2$ film from ambience and substrate, leading to reliable device characteristics with zero persistent photocurrent. On the other hand, they serve as closely separated carrier collection layers, where the separation is governed by the thickness of the $MoS_2$ film. Such small separation between electrodes, which is difficult to achieve in lithography limited planar structures, results in short carrier transit time leading to fast response and high gain, while maintaining a low dark current. This leads to an extremely large $D^*$ of $3.2 \times 10^{14}$ Jones and $R$ of $5.77 \times 10^4$ AW$^{-1}$ observed in the device at $P_{op} = 0.26$ Wm$^{-2}$ and $V_{ext} = -0.5$ V. Further, the asymmetric nature of the device even allows zero external bias operation, under which we obtain a $D^*$ in excess of $10^{14}$ Jones.

## 2. Results and Discussions

Figure 1a-e illustrates the process steps for fabrication of the device. Si wafer coated with 285 nm SiO$_2$ is used as the substrate for the device fabrication. First, using optical lithography, patterns for bottom electrode are realized. This is followed by the deposition and subsequent lift-off of a 20/50/10 nm thick stack of Cu$_2$O/Au/Cr. Cu$_2$O was deposited by reactive ion sputtering of Copper in the presence of Oxygen (0.8 sccm), Nitrogen (5 sccm) and Argon (27 sccm) at 95 W RF power. MoS$_2$ flakes with thickness varying in the range of 10-25 nm were exfoliated micro-mechanically on these electrodes. Using electron beam lithography, PMMA is hardened on the MoS$_2$ flake, leaving the central portion on the flake as a window for the top contact. Using RF sputtering, 20 nm thick transparent metal oxide (Indium Tin Oxide, ITO) is deposited as the top electrode. Finally, 10/50 nm thick Cr/Au contact pads are made using electron beam lithography, followed by electron beam evaporation of metals and lift-off. These devices are Al wire bonded for photocurrent measurements. The optical image of a final fabricated wire bonded photodetector chip is shown in Figure 1f. The cross section of the final device is schematically illustrated in Figure 1g. The corresponding cross section scanning electron micrograph (SEM) is shown in Figure 1h. The optical image of the device and a zoomed scanning electron micrograph are shown in Figure 1i-j. For the control devices, the same process flow is followed except the exfoliation of MoS$_2$ layer in between ITO and Cu$_2$O.

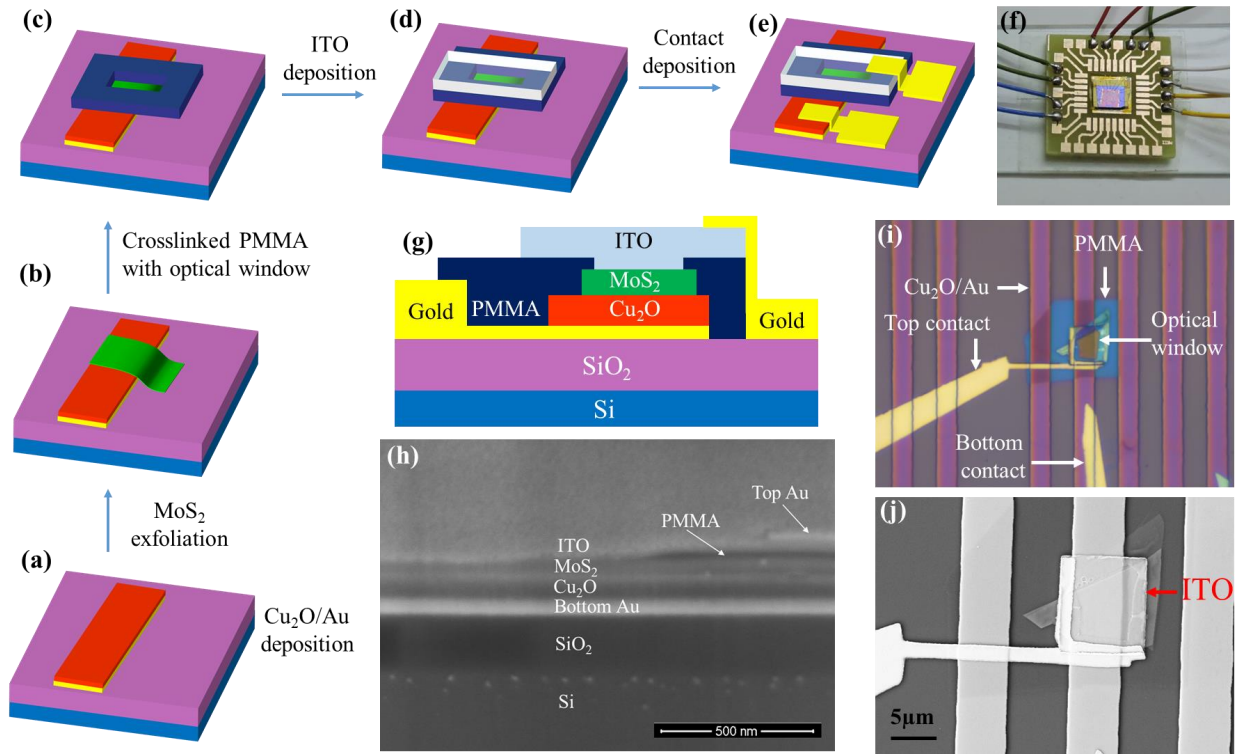

**Figure 1. Fabrication of vertical photodetector.** (a-e) Detailed process flow of the device. (f) An optical image of the wire bonded vertical photodetector chip. (g) Schematic view of cross section of the vertical device. (h) Cross section scanning electron micrograph (SEM) of the device after complete fabrication. (i-j) Optical image and top view SEM of the device after complete fabrication and measurement.

As schematically shown in Figure 2a, the transparency of the top ITO layer plays a key role in determining the efficiency of the photon absorption of the whole device as the active MoS$_2$ layer is buried underneath the ITO layer. Figure 2b illustrates the measured wavelength dependent transmittance of 20 nm thick ITO. In the same plot, we also show that the underneath Cu$_2$O layer exhibits good transparency as well. Considering MoS$_2$ film thickness ($t$) as 20 nm, and taking MoS$_2$ absorption coefficient[15,36] ($\alpha$) as $2 \times 10^5$ cm$^{-1}$, the amount of incoming light absorbed by the MoS$_2$ film is $(1 − e^{−αt}) × 100\% ≈ 33\%$. The fraction of the light which is transmitted through by the MoS$_2$ layer, is reflected back to the active MoS$_2$ layer again by the underneath Au layer through transparent Cu$_2$O, effectively improving the overall light matter interaction length in the MoS$_2$ layer.

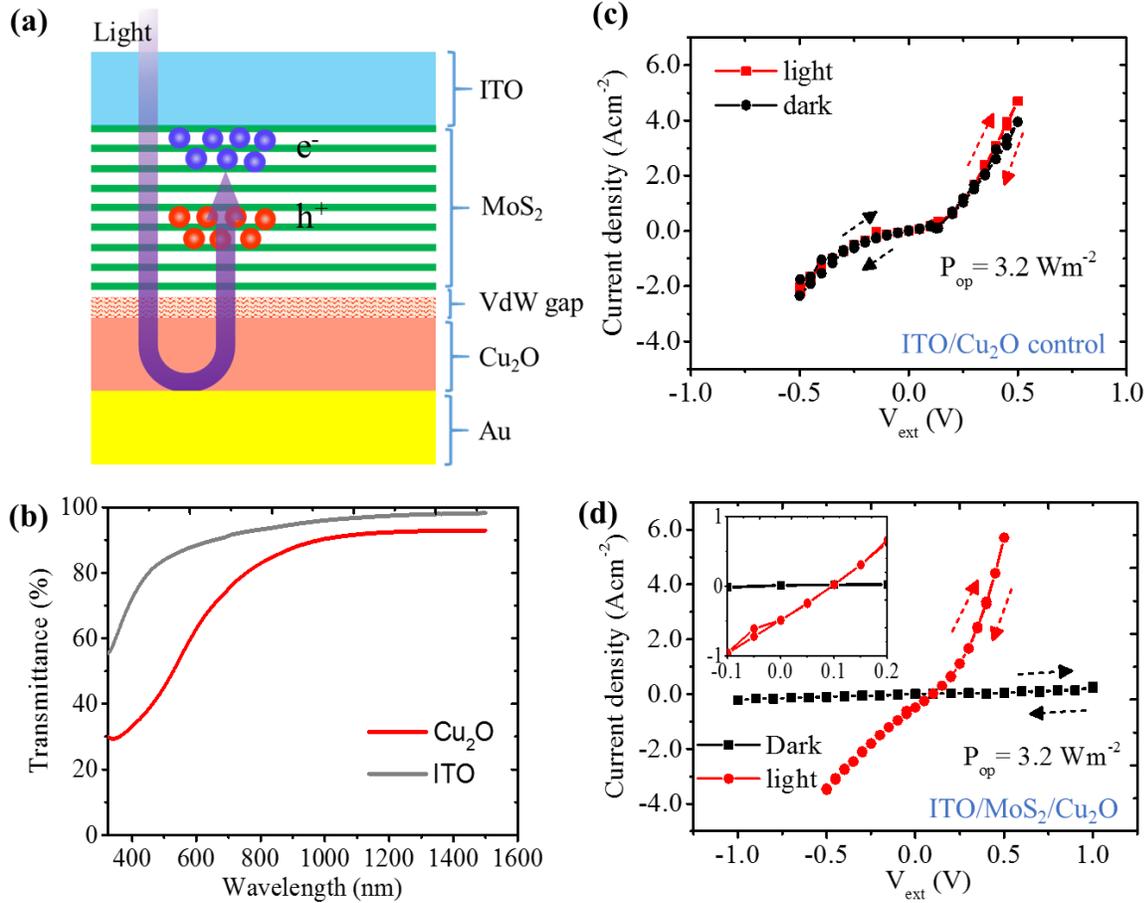

**Figure 2. Photoresponse of the vertical photodetector.** (a) Schematic diagram of the device operation where photons are transmitted through ITO and absorbed by MoS$_2$ layer, followed by separation of photogenerated electrons and holes by strong vertical field. (b) Wavelength dependent transmittance of individual ITO and Cu$_2$O layer. (c) $J − V_{ext}$ characteristics of ITO/Cu$_2$O control device, in light ($\lambda = 600$ nm) and dark conditions, indicating negligible photo response. The arrows indicate voltage sweep direction. (d) $J − V_{ext}$ characteristics in dark and light

($\lambda = 600$ nm) conditions for the vertical device. Forward and reverse sweeps, indicated by arrows, almost coincide. Inset, a zoomed in figure around $V_{ext} = 0$ V, indicating strong photoresponse at zero external bias.

The ITO and Cu$_2$O layers act as heavily doped n-type and p-type semiconductors, respectively, with large conductivities (see Supporting Information S1 for Cu$_2$O transport properties), as tested separately. First, we examine the photoresponse characteristics of the ITO/Cu$_2$O control device, as shown in Figure 2c. Clearly, the current density with light ($J_{ph}$) is almost the same as the dark current density ($J_{dark}$), indicating negligible photoresponse. Also, the current density in forward and reverse sweeps coincide on top of each other. This suggests negligible effect of charge trapping on current voltage characteristics.

Figure 2d illustrates the low dark current and strong photoresponse characteristics of the ITO/MoS$_2$/Cu$_2$O photodetector exhibiting three important features: (i) Negligible hysteresis between forward and reverse sweep is maintained with and without photo excitation, which suggests strong suppression charge trapping effects. This is attributed to the efficient isolation of the MoS$_2$ layer from ambience as well as SiO$_2$ interface, as schematically represented in Figure 3. (ii) The dark current density is effectively suppressed due to the introduction of the few layer MoS$_2$. However, in the presence of light, a strong photocurrent is observed, leading to an on/off ratio of $\approx$ 36 (150) at $V_{ext} = -0.5$ V (+0.5 V). (iii) The device shows pronounced photocurrent at $V_{ext} = 0$, as shown in the inset of Figure 2d. This is attributed to the built-in field induced separation of photo-generated electrons and holes in the MoS$_2$ layer resulting from the asymmetry of the design, as illustrated schematically in the band diagram in Figure 4a.

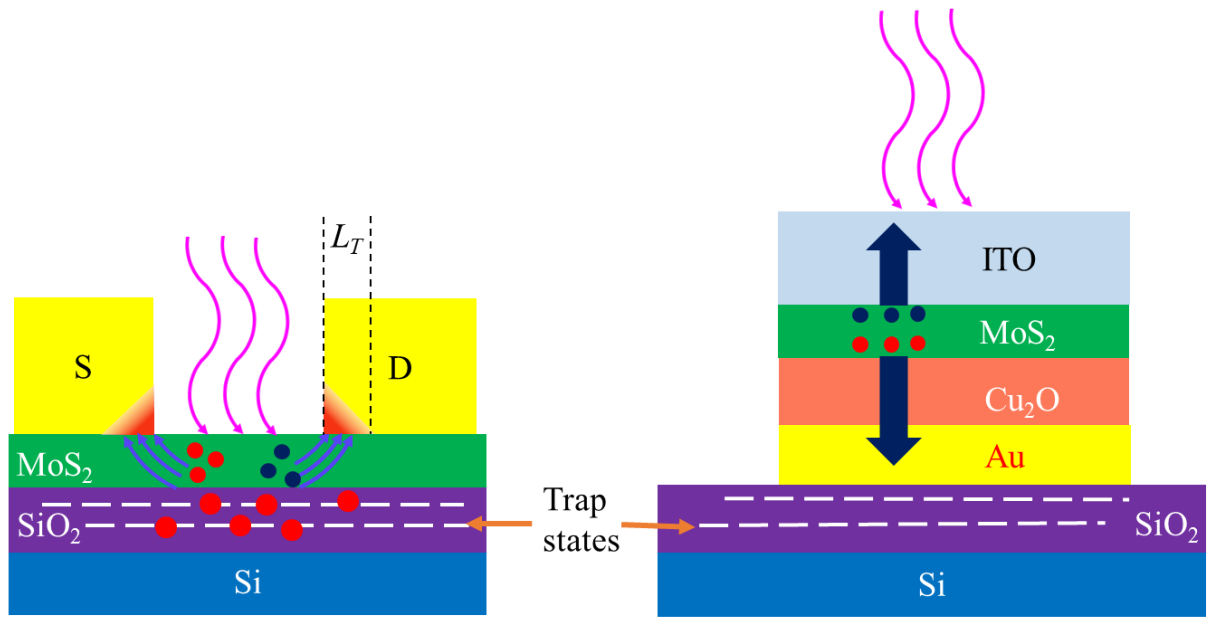

**Figure 3. Isolation from traps and elimination of current crowding in vertical photodetector.** Left panel, carrier collection is bottlenecked by in-plane transport in 2-D material and current crowding at the 2-D/metal interface. The red and blue spheres indicate photo-generated holes and electrons, respectively. The slow traps in $SiO_2$ interact strongly with the active part of the photodetector, efficiently trapping photo-generated carriers. Right panel, photo-generated carriers are collected vertically by ITO and $Cu_2O$ layer. The bottom $Cu_2O$/Au film also completely isolates the device from the traps residing in $SiO_2$ substrate.

While monolayer TMD semiconductors exhibits large exciton binding energy[37,38], the binding energy is expected to be significantly suppressed for multi-layer $MoS_2$. This suppressed excitonic energy helps in efficient separation of the photo-generated electrons and holes. We have further checked that removing the $Cu_2O$ layer (i.e. ITO/ $MoS_2$/Au stack) increases the dark current significantly, while almost entirely suppressing the photocurrent. This is due to large metal induced doping of $MoS_2$[39,40] providing negligible barrier for the electrons.

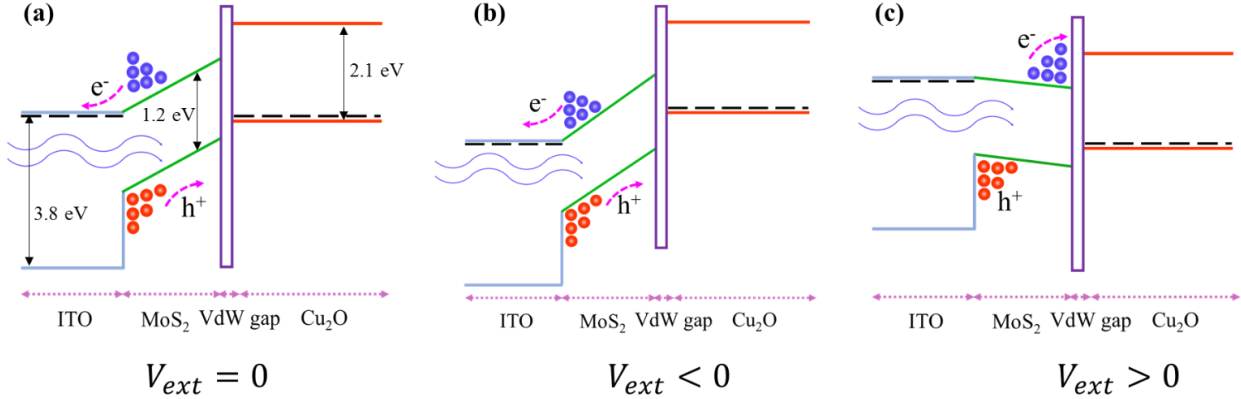

**Figure 4. Band diagram of the device in the vertical direction.** Band diagram (a) in equilibrium, (b) under reverse bias, and (c) under forward bias conditions. In zero bias and reverse bias condition, the electrons (indicated by blue spheres) are collected efficiently by the ITO layer, while the holes (red spheres) are partially blocked by the VdW gap. In forward bias, electron collection efficiency reduces by the thermal barrier at the MoS$_2$/Cu$_2$O junction, while the holes are completely blocked by the large ITO hole barrier, leading to an improved overall gain.

The estimated $R$ and $D^*$ of the device as a function of $V_{ext}$ is plotted in Figure 5a. $R$ is calculated as $R = \frac{\Delta J_{ph}}{P_{op}} = \frac{J_{ph} - J_{dark}}{P_{op}}$. By noting that $R = G\frac{q\eta\lambda}{hc} = G\eta\frac{\lambda \ (nm)}{1243}$ where $G$ is gain, $\lambda$ is excitation wavelength and $\eta$ is the external quantum efficiency, the large $R$ observed suggests that the device has a large internal gain. The origin of the gain mechanism in this architecture can be explained by the asymmetric band offsets at the ITO/ MoS$_2$ and MoS$_2$/Cu$_2$O interfaces as shown in Figure 4. This asymmetry of the device is further enhanced by the way in which both ITO/ MoS$_2$ and Cu$_2$O/ MoS$_2$ contacts are realized. In this case, MoS$_2$ is exfoliated mechanically on Cu$_2$O bottom contact. This leaves a van der Waals (VdW) gap between the two. But the top ITO contact is realized by the RF sputtering on MoS$_2$ and hence this makes an intimate contact. Consequently, at $V_{ext} \leq 0$, the photo-generated electrons are collected efficiently by the ITO layer, while the hole collection is hindered by the VdW gap.

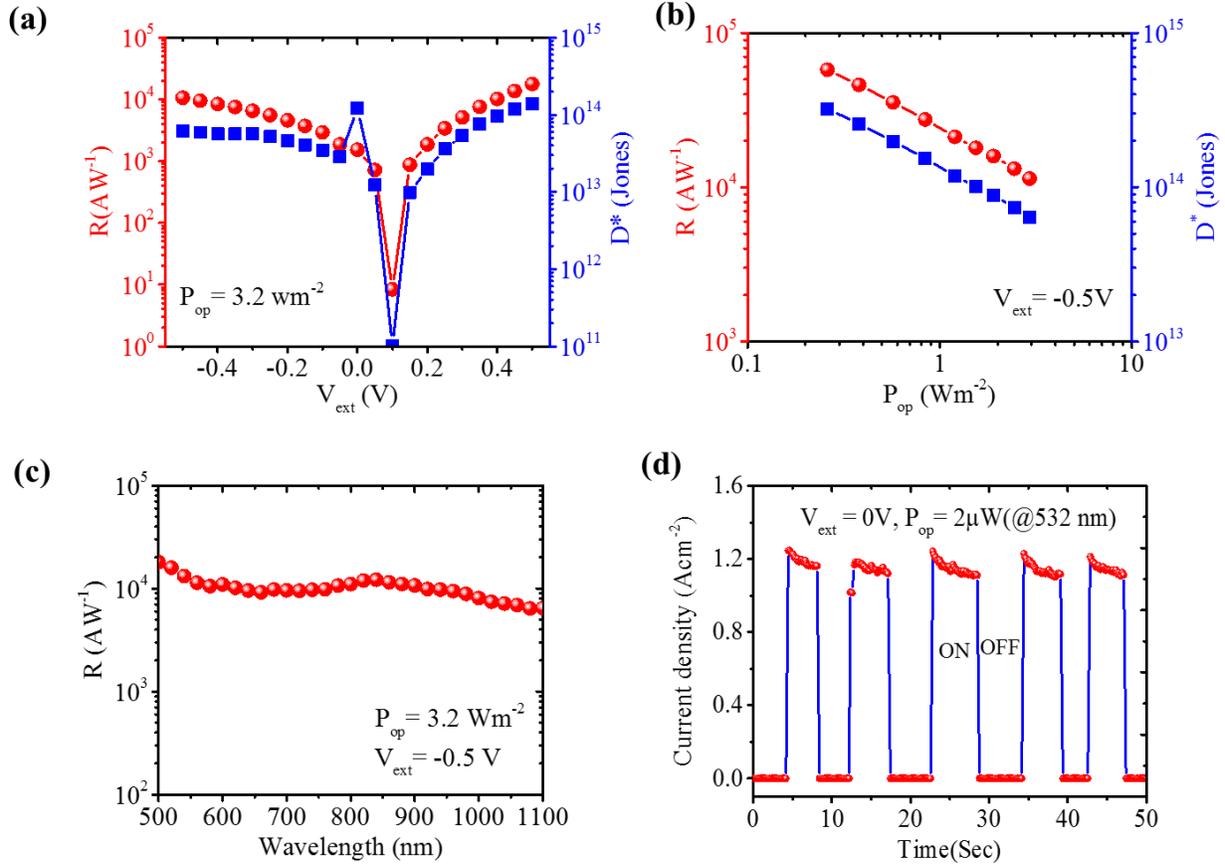

**Figure 5. Photodetection performance.** (a) Responsivity on the left panel and specific detectivity on the right panel, as a function of $V_{ext}$. The spike in $D^*$ at $V_{ext} = 0$ V results from suppression of dark current density. (b) $R$ and $D^*$ as a function of incident optical power density. (c) Wavelength dependent responsivity indicating almost flat response over a large range of wavelengths. (d) Transient response of the device with a 532-nm laser being turned on and off. Fast rise and fall suggest no trap assisted slow photocurrent build-up and persistent photocurrent.

The asymmetric nature of the two contact interfaces is amplified by the difference in valley degeneracy ($g_v$) of the conduction and valence band edges in multi-layer MoS$_2$. The conduction band edge of multilayer MoS$_2$ originates from Q valley (along Γ point to K point) with $g_v = 6$, while the valence band edge is at the Γ point[41] with $g_v = 1$. Since larger valley degeneracy reduces contact resistance[40], electrons see a less resistance path than holes.

Such asymmetric electron and hole collection leads to successive reinjection of electrons until the hole either recombines with an electron or tunnels out of the MoS$_2$ layer through the VdW gap. The gain is given by $G = \frac{\tau_h}{\tau_{tr,e}}$, where $\tau_h$ is the hole lifetime and $\tau_{tr,e} = \frac{t}{\mu_e E}$ is the transit time of an electron. Here $t$ is the MoS$_2$ layer thickness, $\mu_e$ is out of plane electron mobility and $E$ is the effective out-of-plane electric field inside MoS$_2$. The vertical nature of the device inherently has small $t$ and large $E$, resulting in short electron transit time. This leads to a large gain, suggested by the large responsivity obtained from these devices (Figure 5a).

The larger $R$ in forward bias ($V_{ext} > 0$) observed in Figure 5a stems from a higher gain compared with the reverse bias condition ($V_{ext} < 0$). This can be understood from the band diagrams in Figure 4b-c. Under forward bias, the electrons need to thermionically overcome the MoS$_2$/ Cu$_2$O barrier, resulting in less efficient electron collection than in reverse bias. However, the holes are now completely blocked by large hole barrier provided by ITO, significantly enhancing $\tau_h$. Overall, this results in an enhanced gain compared with the reverse biased condition.

An important factor for large photoresponse is efficient collection of photo-generated carriers. One of the bottlenecks in planar two-dimensional photodetectors is the inefficient carrier collection by the metal contacts, which leads to a large series resistance. This arises due to current crowding effect at the metal/2-D interface as schematically illustrated in the left panel of Figure 3. The effective area in which carriers are collected is $W \times L_T$ where $W$ is the channel width and $L_T$ is the transfer length[40,42] of that metal/2-D semiconductor interface. Typically, $L_T < 100$ nm for metal/MoS$_2$ interfaces[42], hence the effective contact area can be much smaller than the physical contact area between metal and MoS$_2$. The vertical transport of the present design, on the other hand, avoids such current crowding by pushing the current crowding region from MoS$_2$ layer to the highly conductive (and practically equipotential) electrode[43]. Effectively, this allows for carrier collection across the entire vertical device footprint, as schematically shown in the right panel of Figure 3.

Assuming the shot noise from the dark current as the primary contributor to the total noise, the $D^*$ has been calculated as[44] $D^* = \frac{R}{\sqrt{2qJ_{dark}}}$, where $q$ is absolute value of electron charge. The suppression of dark current density, coupled with large gain, helps us to obtain a very impressive $D^*$ in the device. In particular, the obtained $D^*$ values are $1.4 \times 10^{14}$ Jones with $V_{ext} = 0.5$ V, at an optical power density of 3.2 Wm$^{-2}$. Also, the device is able to achieve a $D^*$ of $1.2 \times 10^{14}$ Jones without any external bias due to suppression of the dark current.

The obtained values of $R$ and $D^*$ are plotted in Figure 5b as a function incident optical power density. We observe a linearly decreasing $R$ and $D^*$ with an increase in $P_{op}$ in the chosen range of $P_{op}$. Such a decrease can be attributed to the larger accumulation of hole density at higher $P_{op}$, resulting in (i) reduced vertical electric field in the MoS$_2$ layer, and (ii) enhanced recombination of photo-generated electron-hole pairs. At the lowest applied power density of $0.26$ Wm$^{-2}$, we achieved a responsivity, $R = 5.77 \times 10^4$ A/W and a specific detectivity, $D^* = 3.2 \times 10^{14}$ Jones at $V_{ext} = -0.5$ V. In Figure 5c, we show that the strong photoresponse of the device is maintained from $\lambda = 500$ nm to $\lambda = 1100$ nm, exhibiting almost flat wavelength response.

As noted earlier, one of the drawbacks of planar 2-D photodetectors is slow response and persistent photocurrent when the light source is turned off. This results from oxide substrate induced traps being filled by carriers, which in turn reduces the potential barrier that the carriers encounter at the source edge[45]. The ITO/MoS$_2$/Cu$_2$O structure used in this work eliminates such trap assisted persistent photocurrent by isolating SiO$_2$ charge traps, as schematically illustrated in Figure 3. This is suggested by the absence of hysteresis in photocurrent (Figure 2d) as well as fast transient response in Figure 5d. We also note that the typically observed slow building up of the photocurrent in planar 2-D devices during light turn-on is effectively eliminated in the vertical device. The device was tested till 70 ms rise time, which was only limited by the measurement setup. In addition, since the active MoS$_2$ layer is completely encapsulated by top ITO and bottom Cu$_2$O/Au protective layers, we were able to repeat the measurement over many cycles, significantly improving the variability and degradation of performance due to ambience effects. Transient response to 532-nm and 785-nm lasers of devices from a different run are shown in Supporting Information S2. For comparison, we provide the transient response of SiO$_2$ substrate supported MoS$_2$ monolayer and multi-layer lateral devices in Supporting Information S3.

To benchmark the results obtained with existing reports from different photodetector technologies, one key challenge is to compare data from different measurement conditions such as varying optical power density and external bias. To have a fair comparison, in Figure 6, we populate $R$ and $D^*$ versus $P_{op}$ from literature and compare with results obtained from this work at different $V_{ext}$. Keeping practical applications in view, only those data points are considered where the device response time is less than 1 second. The solid line in Figure 6a-b indicates a linear regression trendline of the reports from literature. For reference, state of the art Si and InGaAs photodetector data[44] are also shown in the $D^*$ plot in Figure 6b. The obtained results from the present work clearly look promising over different technologies, providing superior $D^*$ at a given optical power density.

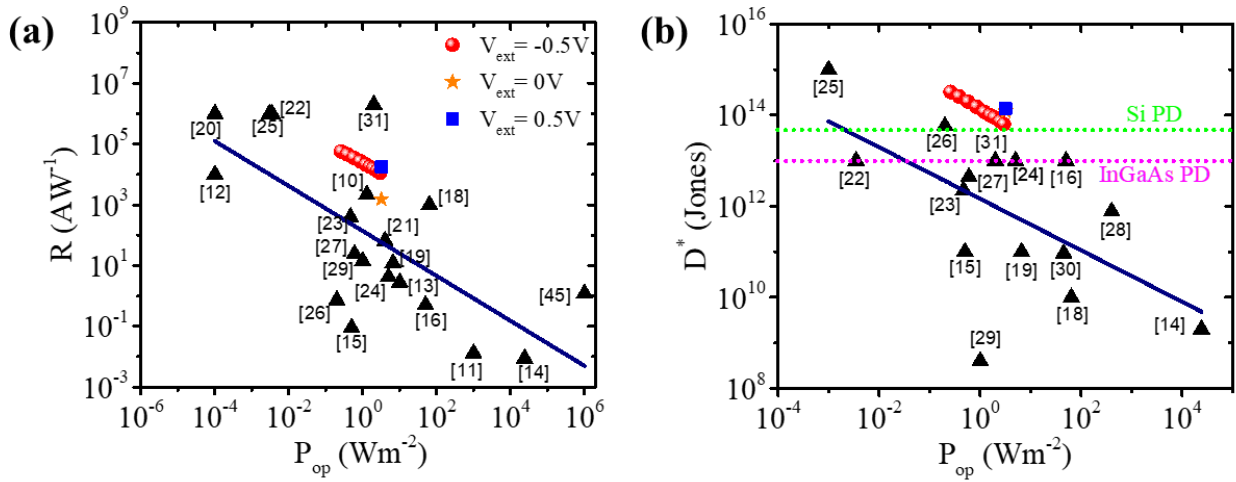

**Figure 6. Benchmarking of device performance.** (a) Responsivity, and (b) specific detectivity, as a function of optical power density. The black triangles indicate data from literature with response time less than 1 second. The blue solid lines indicate linear regression trend line of the reported data in literature. The red solid spheres, orange star, and blue square are data from this work with $V_{ext} = -0.5$ V, 0 V and +0.5 V, respectively. The Si and InGaAs photodetector performances are shown in (b) for comparison.

## 3. Conclusion

In conclusion, we demonstrated a 2-D material based vertical photodetector where reliable and almost trap free operation is achieved by isolating the active portion of the device from the substrate and ambience by the encapsulation of closely spaced, highly conductive, transparent vertical carrier collection layers. The small transit time and low dark current of the structure allow us to achieve extraordinarily high specific detectivity in excess of $10^{14}$ Jones at moderate optical power density, even without any external bias. The proposed design is a new class of layered material based highly sensitive, inexpensive photodetector which will open up a new dimension in fast and ultra-low intensity photodetection applications.

ACKNOWLEDGMENT

The authors also acknowledge the support from Veerapandi, Reshma, and Suma B. N for device fabrication and characterization. K.M. would like to acknowledge support of a start-up grant from IISc, Bangalore; the support of a grant under Indian Space Research Organization (ISRO); grants under Ramanujan fellowship, Early Career Award, and Nano Mission under Department of Science and Technology (DST), Government of India; and a young faculty grant from Robert Bosch Center for Cyber Physical System.

# Supporting Information:

**S1. Two probe current-voltage characteristics of Cu$_2$O**

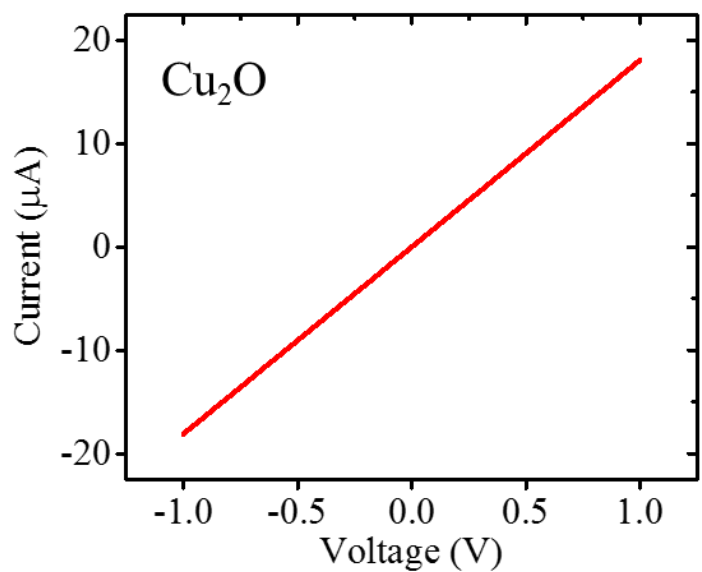

**Figure S1.** Two probe current-voltage characteristics of Cu$_2$O, where probes are around 1 mm away, showing high conductivity and excellent ohmic behavior.

## S2. Transient photoresponse from a different run

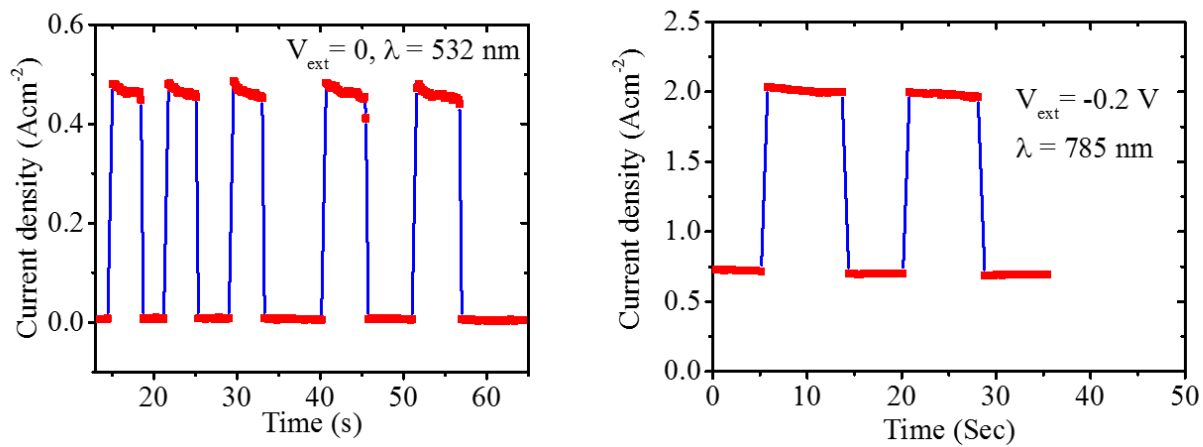

**Figure S2.** Transient response from a different run, with excitation from 532-nm and 785-nm lasers being turned on and off.

## S3. Transient photoresponse of SiO$_2$ substrate supported MoS$_2$ lateral device

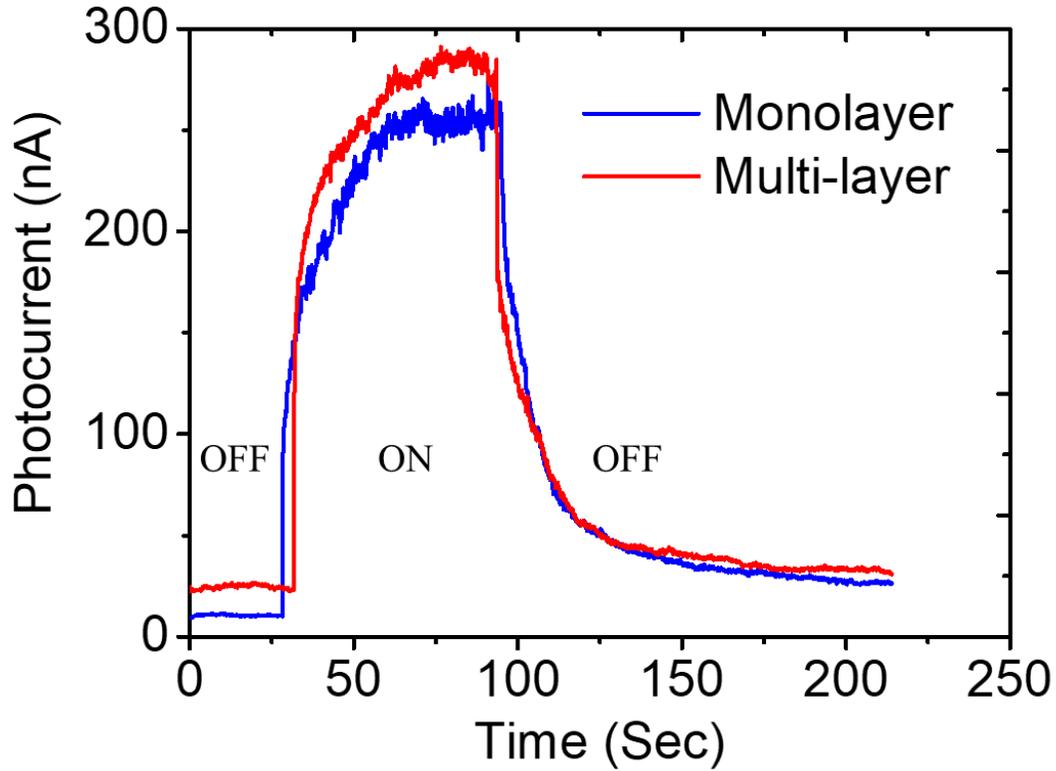

**Figure S3.** Transient photoresponse of monolayer and multi-layer MoS$_2$ lateral photodetector, where the MoS2 film is supported by SiO$_2$ substrate. Both the rise and fall times are found to be much larger than vertical photodetector. ON and OFF indicate regions where the light is turned on and off.